\documentclass{aa}
\usepackage{amsmath}
\usepackage{graphicx}
\usepackage[scriptsize]{subfigure}
\usepackage{lipsum}
\usepackage{natbib}

\title{A new approach to multi-frequency synthesis \\ in radio interferometry}

\author{H. Junklewitz\inst{\ref{inst1},\ref{inst2}} \and M. R. Bell\inst{\ref{inst1}} \and T. En\ss lin\inst{\ref{inst1},\ref{inst2}}} 

\institute{Max-Planck Institut f\"ur Astrophysik (Karl-Schwarzschild-Str. 1, D-85748 Garching, Germany)\label{inst1} \and Ludwig-Maximilians-Universit\"at M\"unchen (Geschwister-Scholl-Platz 1, D-80539  M\"unchen, Germany)\label{inst2}}

\abstract{We present a new approach to multi-frequency synthesis in radio astronomy. Using Bayesian inference techniques, the new technique estimates the sky brightness and the spectral index simultaneously. In principle, the bandwidth of a wide-band observation can be fully exploited for sensitivity and resolution, currently only limited by higher order effects like spectral curvature. Employing this new approach, we further present a multi-frequency extension to the imaging algorithm \textsc{resolve}. In simulations, this new algorithm outperforms current multi-frequency imaging techniques like MS-MF-CLEAN.}
\begin{document}

\maketitle

\section{Introduction}

In radio astronomy, multi-frequency observations are widely used for many different purposes. Examples include the investigation of spectral line emission, analyzing continuous synchrotron spectra of radio sources, the determination of Faraday rotation in polarization imaging, correlating brightness structures at largely different wavelengths or improving the sensitivity and coverage of an interferometer without introducing more antennas into the array \citep[for a review see e.g.][]{WhiteBook}. Most of these only become possible with observations that span over many frequencies and data analysis methods that can handle this kind of observations. 

For the remainder of this work, we focus entirely on multi-wavelength radio continuum studies in total intensity. Neither spectral line observations, polarization imaging, nor wavelengths studies beyond pure radio observations will be a direct topic. In Sec.~\ref{Sec: Conclusion}, we will comment on possible extensions of the presented work into other domains of multi-frequency radio astronomy.

Historically in radio interferometry, the term \textit{multi-frequency synthesis} is mainly used to denote techniques that focus on the direct combination of single-instrument observations at different frequencies to improve the resolution and sensitivity of a radio interferometer (see \citet{MFS} and references therein). For this purpose, a number of methods have been devised, most notably double deconvolution \citep{MFS} and multi-frequency CLEAN \citep{MFClean}. These methods usually work by Taylor-expanding a spectral model function around a reference frequency. We will comment on these methods in more detail in Sec.~\ref{SSec: MFSalpha}.  

Conversely, investigating the spectral behavior of a radio synchrotron source on its own usually is achieved through totally independent standard imaging of the surface brightness in the sky at a number of frequencies. Resolution is kept uniform over all frequencies to correctly recombine the images at individual frequencies. Subsequently, the spectral parameters are determined by fitting a function through all the single frequency images, usually assuming a power-law shaped spectral evolution. 

Only recently, implementations of MF-CLEAN also make it possible to constrain the spectral properties \citep[see][]{CASA}, effectively starting to merge both multi-frequency high resolution imaging and spectral analysis into one algorithm. It is this combined notion, how we like to understand and use the term multi-frequency synthesis.

Further development of multi-frequency imaging techniques is paramount for being able to fully exploit the data from the new generation of radio telescopes, such as the upgraded VLA, LOFAR, the SKA pathfinder missions or ultimately the SKA itself \citep[see e.g.][]{TelescopeReview}. Their unprecedented, broadband frequency coverages of many GHz of possible bandwidths and previously unknown frequency regimes offer many advances in astrophysical and cosmological sciences. But at the same time, they are a challenge for current data analysis methods (see Sec.~\ref{SSec: MFSalpha}). One part of the challenge rises from the fact that current algorithms might not meet the expected sensitivity or fidelity because previously acceptable models and approximations break down with the quality and quantity of data now available. The other side is that the huge quantities in which new data sets come, represent a huge computational challenge on their own, regardless of the algorithms used. It should be emphasized that this study focuses entirely on the first part of the challenge (for this topic, see also Sec.~\ref{Sec: Conclusion}).

In this paper, we present a new approach that uses Bayesian statistical inference techniques to combine a \textit{simultaneous} statistical estimation of the sky brightness and the spectral behavior with the benefits in resolution and sensitivity of classical multi-frequency synthesis. Conceptually, this new approach has a number of advantages over standard methods. In principle, the full bandwidth of the observation is used for maximum theoretical sensitivity. No subsequent imaging at all single frequencies is employed and, thereby, less reconstruction artifacts are introduced and no artificial downgrading of resolution is necessary. Our model for the spectral behavior is not approximated through Taylor-expansions around a frequency, and in the moment only is limited by possible higher order effects like spectral curvature. In this way, our approach is conceptually very similar to the new method of \textit{Faraday Synthesis} \citep{FS} in polarization imaging (see Sec.~\ref{SSec: MFResolve} for more details). Furthermore, the spatial correlation of the parameters describing the spectral behavior is used to constrain and improve the estimation of spectral properties, and can further be viewed as a new scientific result on its own. Finally, it is possible to approximate the statistical uncertainty of the spectral index estimate in addition to that of the total intensity.

To demonstrate the viability of our new approach, we present a multi-frequency extension to the radio extended emission imager \textsc{resolve} \citep{resolve} as an alternative to multi-scale-multi-frequency CLEAN \citep{MSMFClean}, the standard method for wide-band multi-frequency synthesis of extended radio sources.

For our derivations, we will often refer to the work presented in \citet{resolve}, henceforth Res2013.

\section{Theory \& Algorithm}

In this section, we first briefly review the current status of multi-frequency synthesis techniques (Sec.~\ref{SSec: MFSalpha}), then develop the new approach (Sec.~\ref{SSec: MFResolve}) based on \textit{information field theory} \citep{IFT} and the inference framework presented in Res2013. We further present a multi-frequency-extension to the imaging algorithm \textsc{resolve} (see Res2013) using the developed framework (Sec.~\ref{SSec: MFResolve}).

\subsection{Multi-frequency aperture synthesis and spectral index reconstruction}\label{SSec: MFSalpha}

Aperture synthesis is the technique of connecting an array of telescopes in such a way that we can effectively synthesize a combined instrument with a much larger aperture and therefore resolution \citep{Ryle, 1986isra.book.....T}. It can be shown that for observations at a single frequency $\nu_0$, and under the assumption of measuring the sky as flat in a plane tangent to the phase center of the observation, such a radio interferometer approximatively takes incomplete samples of the Fourier transformed brightness distribution in the sky \citep{1986isra.book.....T}:

\begin{align}
V(u,v, \nu_0) &\approx W(u,v,\nu_0) \int \mathrm{d}l \ \mathrm{d}m \ I_{0}(l,m,\nu_0) \ \mathrm{e}^{-2 \pi i \left(ul + vm\right) }. \label{Eq: basicequation}
\end{align}
For our purposes, working under this assumption suffices for the analysis undertaken in this paper. For a detailed derivation of (\ref{Eq: basicequation}) consider \citep{1986isra.book.....T}.
The quantity $V(u,v)$ is called the \textit{visibility} function. The coordinates $u$ and $v$ are vector components describing the distance between a pair of antennas in an interferometric array, where this distance is usually referred to as a \textit{baseline}. They are given in numbers of wavelengths, with $u$ and $v$ usually parallel to geographic east-west and north-south, respectively. The coordinates $l$ and $m$ are a measure of the angular distance from the phase center along axes parallel to $u$ and $v$, respectively. $W(u,v)$ is a sampling function defined by the layout of the interferometric array. It is zero throughout most of the $u,v$-space, apart from where measurements have been made where it is taken to be unity.

The visibility function is what our instrument measures, but we are actually interested in the brightness distribution of the source in the sky. Unfortunately, an inversion of (\ref{Eq: basicequation}) gives us not the true brightness distribution, but its convolution with the inverse Fourier transform of the sampling function, better known as the \textit{dirty beam} $I_{db} = \mathcal{F}^{-1}W$:

\begin{equation}
I_{\mathrm{D}} = \mathcal{F}^{-1}V =  \mathcal{F}^{-1}W \mathcal{F}I = I_{db} \ast I \label{Eq: invprob}.
\end{equation}
Here, we have introduced a symbolic Fourier operator $\mathcal{F}_{kx} = \exp(-i(ul + vm))$ with $x=(l,m)$ and $k=(u,v)$, the common notation $I_{\mathrm{D}}$, \textit{dirty image}, for the simple Fourier inversion of the visibilities, and the symbol $\ast$ to denote a convolution operation.

For imaging at a single frequency, one usually proceeds using a deconvolution algorithm aiming to solve (\ref{Eq: invprob}) approximatively. The most common algorithm is CLEAN \citep{1974A&AS...15..417H}, or one of its many variants \citep{ClarkClean,CS-Clean,MFClean,MSClean,MSMFClean}, which basically model the sky brightness to be a collection of delta peak point sources that are iteratively assigned by searching for the peak values in the dirty image. But alternatives exist, especially for imaging extended emission, like the Maximum Entropy Method \citep{MEM}, Adaptive Scale Pixel decomposition \citep{asp} or the recently published Bayesian extended emission imager \textsc{resolve} (Res2013), which will be used in this paper in an expanded multi-frequency version (see Secs.~\ref{SSec: MFResolve} and \ref{Sec: Tests}).

Originally, radio telescopes observed with relatively narrow bandwidths and at only a few different frequencies. The standard approach for imaging the spectral properties of a source since then is to take single frequency observations, or observations averaged over close channels for higher sensitivity, image them separately, and then fit a spectral model to the observations. Since continuous synchrotron emission is known to show a power-law spectrum \citep{RL}, for many astrophysical purposes such a model is sufficient:

\begin{equation}
I(l,m,\nu) = I_{0}(l,m,\nu_0) \ \left(\frac{\nu}{\nu_0}\right)^{-\alpha}. \label{eq:basicspecdep} 
\end{equation}

Sometimes, higher order spectral deviations are modeled with a second term in the exponential of (\ref{eq:basicspecdep}):

\begin{equation}
I(l,m,\nu) = I_{0}(l,m,\nu_0) \ \left(\frac{\nu}{\nu_0}\right)^{\left(-\alpha + \log{\nu/\nu_0} \ \beta\right)}, \label{eq:speccurve} 
\end{equation}
effectively enhancing the linear function in $\log{I} \ \mathrm{vs.} \ \log{\nu}$ space (\ref{eq:basicspecdep}) to a quadratic polynomial. This model is referred to as spectral curvature. 

Simultaneous multi-frequency imaging at different frequencies has been introduced to radio astronomy upon the realization that observations at many frequencies, if stacked together appropriately, can improve the sampling in the $uv$-plane (see \citet{MFS} and references therein). That way, the sensitivity of a radio interferometric observation can be enhanced considerably. Because the $uv$-coordinates are measured in numbers of wavelengths, the same interferometer samples different parts of the Fourier space at different frequencies. An interferometer includes $N(N-1) / 2$ baselines and therefore $uv$ points, where $N$ is the number of antennas. In theory, observing at a number of frequencies $N_f$ enhances the measured baselines to $N_f N(N-1) / 2$, which is the equivalent effect of introducing roughly $\sqrt{N_f}$ extra antennas \citep{MFS}. 

This approach was first developed by \citet{MFS} for mid-sized fractional bandwidths of around $\pm 10 \%$, together with the method of double deconvolution to mitigate spectral errors when using CLEAN for a spectrally combined data set. Later, double deconvolution was developed into the more advanced multi-frequency CLEAN (MF-CLEAN) \citep{MFClean} and multi-frequency multi-scale CLEAN (MF-MS-CLEAN) \citep{MSMFClean}. 

All these methods assume the spectral dependence to be a power-law, like Eqs.~(\ref{eq:basicspecdep}) or (\ref{eq:speccurve}) and propose to approximate it during the CLEAN-like-deconvolution process with a Taylor-expansion. To our knowledge, current implementations use a few terms of a Taylor-expansion in $\nu$ or $\log{\nu}$ around a reference frequency $\nu_0$. In \citet{MSMFClean} it is discussed that a direct decomposition into one or two terms of a polynomial in $\log{I} \ \mathrm{vs.} \ \log{\nu}$ would actually be the most accurate representation\footnote{Actually, for the spectral model  (\ref{eq:speccurve}), this would be no approximation at all since it simply is a quadratic polynomial in $\log{I} \ \mathrm{vs.} \ \log{\nu}$ space.}, but for current implementations this is discarded by the authors because of numerical instabilities. 

During the CLEAN deconvolution process, the iteratively updated sky model is used to also update the coefficients of the spectral Taylor expansion. In this way, spectral index or even spectral curvature can be constrained by these coefficients. Care must be taken, since this expansion is usually stopped after a few terms and might not be valid over large bandwidths. Because of this, the only implementation of MF-MS CLEAN known to the authors to date, within the radio astronomical software package CASA \citep{CASA}, is considered to be experimental and on a shared-risk basis with regard to the spectral reconstructions with a higher number of terms.

For modern wide-band data sets from the new generation of instruments, spanning several GHz of bandwidths, this means that probably even more emphasis must be put onto the handling of the spectral effects, either by invoking higher order terms in the expansions \citep[see][]{MFClean,MSMFClean} or by shifting to a different approach, where the sky brightness and its spectral properties are fully considered simultaneously, and estimated to fit the entire data using some global minimization function. This last approach was actually mentioned by \citet{MFS}, but found to be unnecessarily complicated and computationally expensive for the typically modest fractional bandwidths at the time. 

We will now consider the latter approach and present a statistical solution. 

\subsection{A multi-frequency extension to the algorithm \textsc{resolve}} \label{SSec: MFResolve}

In the course of this section, we often refer to the detailed derivations layed out in Sec.~2 and App.~1 of Res2013 that form the basis from which we derive our multi-frequency algorithm.

We start by summarizing (\ref{Eq: basicequation}) and (\ref{eq:basicspecdep}) into a multi-frequency measurement model:

\begin{equation}
d(k,\nu) = W(k,\nu) \int dx \ \mathcal{F}(k,x) I(x,\nu) + n(k,\nu) \label{eq:basicmfs}
\end{equation}
where again $x=(l,m)$ and $k=(u,v)$, the term $n(k,\nu)$ introduces measurement noise, the data $d(k,\nu)$ has been introduced, which is basically the visibility function with measurement noise,  and the spectral dependence of the source $I(x,\nu)$ is kept general for the moment.

In order to simplify notation and to identify (\ref{eq:basicmfs}) as a general inverse inference problem as analyzed in Res2013, we henceforth drop all explicit dependence on $k$ and $x$ and combine all known instrumental effects into a response function $R_\nu:= W(k,\nu) \mathcal{F}$, leading to

\begin{equation}
d_\nu = R_\nu I_\nu + n_\nu. \label{eq:datamodel}
\end{equation}
There are many instrumental effects beyond the sampling in the $uv$-plane given by $W(k,\nu)$ like an antenna sensitivity pattern or an direction dependent, variable sampling. For this work, we stay with the basic definition $R_\nu:= W(k,\nu) \mathcal{F}$ and refer the reader to Res2013 concerning the possibility of including other effects within this framework. We just emphasize that, without loss of generality, most of these effects can be in principle included\footnote{In principle this approach can be extended into a full RIME (Radio Interferometer Measurement Equation), considering all Stokes parameters and instrumental gains \citep{SmirnovI,SmirnovII}.}. We also assume the instrument to be fully calibrated, and thus the response $R_\nu$ to be known.  

In accordance with standard radio interferometric literature \citep{1986isra.book.....T}, we assume Gaussian noise statistics, mainly induced by the antenna electronics and independent between measurements at different frequencies and time steps of the observation \citep{1986isra.book.....T}. Henceforth, the noise $n_\nu$ will be assumed to be drawn from a multivariate, zero mean Gaussian distribution of dimension $n_d$:

\begin{align}
\mathcal{P}(n) &= \mathcal{G}(n,N) \notag\\
		 &:= \frac{1}{\mathrm{det}(2 \pi N_\nu)^{1/2}} \ \exp\left(-\frac{1}{2} n_\nu^{\dagger} N_\nu^{-1} n_\nu \right). 
\end{align}

Solving (\ref{eq:datamodel}) exactly for $I_\nu$ is not possible, since all information is lost on the Fourier modes not sampled by $R_\nu$. This is just a different way of stating that a direct Fourier inversion of (\ref{Eq: basicequation}) yields the dirty image and not an exact representation of the sky brightness. 

Instead, the aim is to find a statistical estimate for the \textit{most probable} sky brightness signal, given all observational and noise constraints. In Res2013, it was shown in detail how this can be done in a Bayesian statistical framework for a radio astronomical data model like (\ref{Eq: basicequation}). We briefly repeat the main points, and otherwise refer the reader to Res2013.

To find an optimal statistical estimate for the sky brightness signal $I_\nu$, we regard it as a random field with certain \textit{a priori} statistical properties expressed in the prior distribution $P(I)$, but fully constrained by the data through the statistics of the \textit{likelihood} distribution $P(d|I)$. The likelihood distribution summarizes how the data are obtained with a measurement of the true sky brightness signal, and for our problem can be expressed as

\begin{align}
\mathcal{P}(d|I) = \mathcal{G}(d_\nu-R_\nu I_\nu,N_\nu), \label{likeli}
\end{align}
which is a Gaussian over $d_\nu-R_\nu I_\nu$ with the covariance structure of the uncorrelated noise $N_{kk'\nu} = \delta_{kk'} \sigma_{k\nu}^2$.
 
Prior and likelihood statistics can be combined into the \textit{posterior} distribution $P(I|d) \propto P(d|I)P(I)$ that holds the important information of how much the sky brightness signal is statistically constrained by the data. From there, an estimate for the signal can be obtained by calculating a suitable statistic of the posterior, most prominently its mean or its mode, corresponding to the minimization of different error norm measures between the signal and its estimate (see Res2013; or \citet{Jaynes}, \citet{Caticha} and \citet{Lemm}, \citet{IFT} for a comprehensive review of Bayesian statistics or inference on fields respectively). 

The exact choice of an appropriate inference algorithm at this stage largely depends on the complexity of the problem (i.e. the posterior). For the problem at hand, since the likelihood is already known (\ref{likeli}), this comes down to the question of the prior statistics of $I_\nu$. 

The most general way conceivable would be \textit{not} to explicitly model the spectral dependence of $I(x,\nu)$ at all, as for instance in (\ref{eq:basicspecdep}) or (\ref{eq:speccurve}). Instead, $I(x,\nu)$ can be interpreted as a three dimensional continuous field and should be inferred as a whole from the entire data. In general, let us assume not only point-like but also extended emission in the sky, and some kind of extended structure in spectral space as well. In such a setting, $I(x,\nu)$ could be set \textit{a priori} as a statistical field with an unknown and probably non-isotropic cross-correlation structure in the combined sky and spectral space. Such a complex, unknown cross-correlation structure at the one hand complicates the problem enormously, but could also be used to guide the reconstruction, if correctly estimated with the field itself. 

A number of statistical methods have already been developed to handle simpler problems of signal reconstructions with unknown but isotropic correlation structure, many of them solving the problem for Gaussian fields using information field theory \citep{EnsFrom,EnsWeig} or Gibbs-Sampling Monte-Carlo methods \citep{WandeltLine1,WandeltLine2}, or recently also for log-normal fields \citep{EnsFrom,EnsWeig, SmP, MaksimMaster, d3po}. Most notably this method was also used to create \textsc{resolve} (Res2013). A full combined spatial and spectral reconstrcution as outlined above, would require substantial further development and is outside of the scope of this work.

We can choose a more direct strategy, still residing within our approach of statistical inference, but fix a spectral model and infer instead the spectral index (or curvature) as a field on its own. For the rest of the paper, we will choose the model (\ref{eq:basicspecdep}) for the simplicity of the approach, and for a functional similarity with the algorithm \textsc{resolve} that makes it very natural to include into a combined method.

\textsc{resolve} works under the assumption that the extended surface brightness at a single frequency is \textit{a priori} assumed as a random field drawn from log-normal statistics (Res2013). For our multi-frequency problem, this basically turns (\ref{eq:datamodel}) into

\begin{align}
d_\nu &= R_\nu I_\nu + n_\nu \notag\\
      &= R_\nu \left[I_{0}(l,m,\nu_0) \ \left(\frac{\nu}{\nu_0}\right)^{-\alpha}\right] + n_\nu \notag\\
      &= R_\nu \left[\rho_0 \ \mathrm{e}^{s(l,m)} \ \left(\frac{\nu}{\nu_0}\right)^{-\alpha}\right] + n_\nu, \label{resolve}
\end{align}
where $s$ is a Gaussian random field (such that the logarithm of $\mathrm{e}^s$ is a Gaussian random field again), and $\rho_0$ is a constant to e.g. normalize the system to the right units. Although, the frequency dependence of $s$ was not explicitly shown in Res2013, the derivation of \textsc{resolve} implicitly assumed the algorithm to work for a single frequency in the way presented here. \textsc{resolve} assumes the spatial correlation of an extended source in the sky to be reflected by the covariance of the Gaussian random field $s$, which is unknown \textit{a priori}, and thus estimated from the data itself together with the sky brightness. The covariance $S$ of a Gaussian field is equivalent to its two-point correlation function $S(l,m) = \langle s(l)s(m)^\dagger \rangle$ and is handled as such by \textsc{resolve} in form of the power spectrum $S(k,k') = \langle s(k) s(k')^\dagger \rangle$, which is the Fourier transformation of the correlation function\footnote{We actually assume the spatial correlation to be \textit{a priori} rotationally and translationally invariant, and thus the power spectrum to be diagonal $S(k,k') = \left<s(k)s(k')^{\dagger}\right> = (2\pi)^{n_{s}} \delta(k - k') P_{s}(|k|)$. However, this does not imply that the correlation structure must be invariant under any transformation \textit{a posteriori} as well. A more detailed discussion can be found in Res2013.}. A deeper analysis of this can be found in Res2013.  

We now make the central assumption that the spectral index $\alpha$ can be modeled \textit{a priori} as a Gaussian random field with its own spatial correlation structure in the sky. At least for an extended source, we have every reason to assume that the spectral index should be a field with spatial extension itself. Observational constraints strongly imply that typical extended radio structures show as well extended and smooth spectral index structures, for instance radio halos and relics of galaxy clusters \citep{ClusterReview}, radio galaxy lobes \citep{SNR_RG_CL}, or supernova remnants \citep{SNRreview}.

In Res2013, we argued extensively that a Gaussian random field would be the ideal choice for a signal prior of an extended field with a priori unknown correlation structure, as long as the field is not assumed to vary strongly on orders of magnitude and not necessarily needs to be positive definite. Both constraints apply very well to known spectral index maps, where variations usually do not reach even one order of magnitude, and nothing prevents the spectral index in principle to change the sign. For details of these arguments see Res2013, we now proceed under this assumption. 

If we rewrite (\ref{eq:basicspecdep}) only slightly,

\begin{align}
I(l,m,\nu) &= I_{0}(l,m,\nu_0) \ \left(\frac{\nu}{\nu_0}\right)^{-\alpha} \notag\\
	   &= I_{0}(l,m,\nu_0) \ \mathrm{e}^{-\ln{\left(\nu/\nu_0\right)} \alpha}, \label{surprise}
\end{align}
it reveals that, if $\alpha$ has a Gaussian prior, (\ref{surprise}) naturally turns into a model for a log-normal prior, only different in shape (and more complicated) from (\ref{resolve}) because of the term $-\ln{\left(\nu/\nu_0\right)}$.

It is important to note that we have not specified $I_{0}(l,m,\nu_0)$ yet. Thus, at this point, our inference approach to multi-frequency synthesis is in principle compatible with any method that reconstructs and deconvolves the surface brightness $I_{0}(l,m,\nu_0)$ at a single reference frequency $\nu$. At least as long as it seems consistent with the source of interest to assume that the spectral index is an extended and spatially correlated field. In an extreme case, like single, unresolved point sources, this method probably will not yield optimal results.

For this paper, we take the choice to combine (\ref{resolve}) and (\ref{surprise}) into one single method, where we assume our double log-normal measurement model to be

\begin{align}
d_\nu = R_\nu \left[\rho_0 \ \mathrm{e}^{s(l,m) - \ln{\left(\nu/\nu_0\right)} \alpha(l,m)}\right] + n_\nu, \label{mfresolve}
\end{align}
with $\rho_0$ again a constant, from now on set to one w.l.o.g., and the signal fields $s$ and $\alpha$ having Gaussian prior distributions $P(s)$ and $P(\alpha)$:

\begin{align}
\mathcal{P}(s)   &= \mathcal{G}(s,S) \notag\\
		 &= \frac{1}{\mathrm{det}(2 \pi S)^{1/2}} \ \mathrm{e}^{-\frac{1}{2} \ s^{\dagger} S^{-1} s}, \\
\mathcal{P}(\alpha)   &= \mathcal{G}(\alpha,A) \notag\\
		 &= \frac{1}{\mathrm{det}(2 \pi A)^{1/2}} \ \mathrm{e}^{-\frac{1}{2} \ \alpha^{\dagger} A^{-1} \alpha}.
\end{align}

We now write down a posterior distribution for each signal field, while the other field (and its covariance) is assumed to be known, held constant and regarded as part of two distinct versions of the response operator $R_\nu$, henceforth called $R_{(\mathrm{s})} = W(k,\nu) \mathcal{F} \left(\mathrm{e}^{-\ln{\left(\nu/\nu_0\right)} \alpha} \ \circ \right)$ and $R_{(\mathrm{\alpha)}} =  W(k,\nu) \mathcal{F} \left(\mathrm{e}^{s(l,m)} \ \circ \right)$ (where the symbol $\circ$ denotes where the field needs to be inserted that the operator acts on):

\begin{align}
&\mathcal{P}(s|d) \propto \mathcal{G}(d-R_{(\mathrm{s})}\mathrm{e}^{s},N) \ \mathcal{G}(s,S), \label{s_posterior} \\
\notag\\
&\mathcal{P}(\alpha|d) \propto \mathcal{G}(d-R_{(\mathrm{\alpha)}}\mathrm{e}^{-\ln{\left(\nu/\nu_0\right)} \alpha},N) \ \mathcal{G}(\alpha,A). \label{a_posterior}
\end{align}

As in Res2013, it is not possible to calculate the posterior mean for either of the two signal fields $s$ or $\alpha$ without invoking complicated or expensive perturbation or sampling methods (see Res2013). We therefore continue to use the procedure already presented there, and calculate the posterior maximum to estimate both signals

\begin{align}
&m_{\mathrm{s}} = \mathrm{argmax}_{s} \mathcal{P}(s|d), \notag\\
\notag\\
&m_{\mathrm{\alpha}} = \mathrm{argmax}_{\alpha} \mathcal{P}(\alpha|d). \label{posmean2map}
\end{align}
In signal inference, this procedure is called Maximum A Posteriori (MAP) \citep{Jaynes}. The resulting fix-point equations from Eqs.~(\ref{posmean2map}) need to be solved numerically using a non-linear optimization scheme. For this, we resort to the same implementations as in Res2013 (see App.~2 therein).

With this choice, we basically extend \textsc{resolve} to a multi-frequency algorithm by integrating a second complete \textsc{resolve} step for the spectral index into the method, and iterating between the statistical estimation of $s$ with its covariance $S$, and $\alpha$ with its covariance $A$. As outlined above, we always hold one of the fields (and their respective covariances) as constant and regard them as part of the response during this process.

It should be emphasized again that the \textsc{resolve} step for the spectral index could in principle be combined with any other method to reconstruct $I_{0}(l,m,\nu_0)$ at a single frequency $\nu_0$.

The exact equations that need to be solved to calculate (\ref{posmean2map}) for either of the two fields and estimate their power spectra (i.e. their correlation structure in form of their Gaussian covariances, see above) are layed out in App.~\ref{App1} and derived rigorously in Res2013.

\section{Tests} \label{Sec: Tests}

We have integrated multi-frequency capability into the existing implementation of \textsc{resolve} and tested the algorithm using simulated data\footnote{To get access to the preliminary code prior to its envisaged public release, please contact henrikju@mpa-garching.mpg.de or ensslin@mpa-garching.mpg.de.}. The code is written in \textsc{Python} using the signal inference library \textsc{NIFTy} \citep{nifty}, for details of the implementation we refer the reader to App.~2 of Res2013.

\begin{figure}[h!]
\centering
  \subfigure[Single frequency $uv$-coverage in units of \# of wavelengths.]{
   \includegraphics[width=0.45\textwidth]{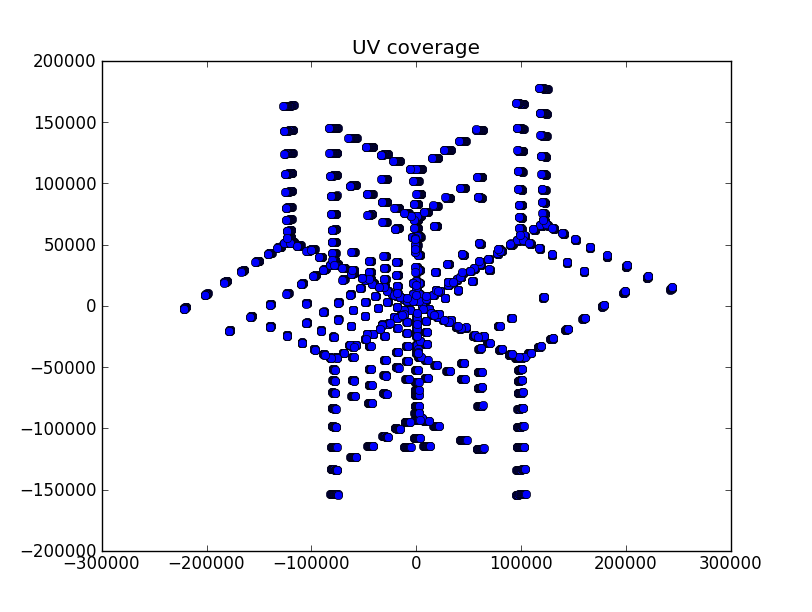}}
  \subfigure[Full multi-frequency $uv$-coverage in units of \# of wavelengths.]{
   \includegraphics[width=0.45\textwidth]{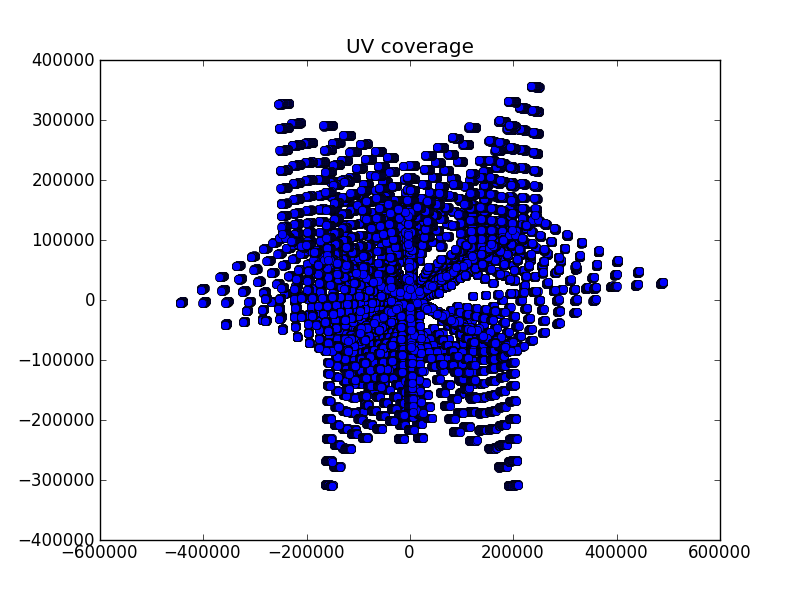}}
  \caption{Single-frequency and full-frequency $uv$-coverage from all frequencies of a simulated 20 minutes snapshot observation in VLA-a configuration. } 
\label{realuvcov}
\end{figure}

As in Res2013, we constructed simulated observations with the tool \textsc{makems}\footnote{See http://www.lofar.org/wiki/lib/exe/fetch.php \\ ?media=software:makems.pdf.} using a realistic $uv$-coverage from a VLA observation in its A-Configuration. The VLA samples the $uv$-plane non-uniformly at irregular intervals, and the response includes thereby a convolutional gridding and degridding operator using a Kaiser-Bessel kernel (for details see App.~2 in Res2013). We simulated observations over a range of 2 GHz, with 20 separate frequency channels. The observations are short snapshots of approximatively 20 minutes per frequency, with a total of 42\,120 visibility measurements at each frequency channel (see Fig. \ref{realuvcov}). This setting leads to an especially sparse sampling of the $uv$-plane at a single frequency, but to a much better coverage for the combined multi-frequency data (see Fig. \ref{realuvcov}). 

For all tests, the signal $s$ is drawn from a Gaussian distribution, finally entering the formalism as an exact log-normal field\footnote{In Res2013 it was demonstrated that \textsc{resolve} also works beyond that on realistic signals drawn from real CLEAN maps.} 

In order to use a spectral index signal with some correlation to the brightness signal, we model the spatial source dependence of the spectral index in an ad hoc fashion. 

For the spectral index signal $\alpha$, we use a sum of two Gaussian fields (see Sec.~\ref{Ssec: mainres}), one being independently drawn, while the second is the surface brightness signal $s$ itself, down-weighted with a suitable factor. This is done in order to introduce a spectral index signal with some correlation to the surface brightness signal, at least in an ad-hoc fashion. Typically, extended sources in radio astrophysics show a spatial cross-correlation between both, which is a result of different physical processes underlying the structure and formation of these sources. A notable example are radio halos and relics in galaxy clusters \citep{ClusterReview}. Of course, our model is only an ad-hoc approximation for the proof of concept undertaken in this work. We also emphasize that such a cross-correlation between $\alpha$ and $s$ is \textit{not} exploited by \textsc{resolve} (see Sec.~\ref{SSec: MFResolve} for an outlook).  

The complex, Gaussian input noise variance in $uv$-space is equal for all visibilities and frequencies. As with single-frequency \textsc{resolve}, the algorithm does not require equal noise variances and can in principle handle varying variances. The noise variance was set to a low\footnote{In comparison to the signal strength in Fourier space, the chosen value ensures a high signal to noise ratio. The unit Jy was used here for convenience. Effectively, it stands for whatever units the simulated signal is interpreted to be given in.} value of $\sigma^2=10^{-3} \mathrm{Jy}^2$.
 
In continuation with Res2013, we use a relative $\mathcal{L}_{2}$ - norm measure of the difference in the signal to the reconstruction to measure the accuracy of the estimate of both brightness and spectral reconstructions:

\begin{align}
\delta_s = \sqrt{\left(\frac{\sum\left(\mathrm{e}^s - \mathrm{e}^m\right)^2}{\sum \left(\mathrm{e}^{s}\right)^2}\right)}, \\
\delta_\alpha = \sqrt{\left(\frac{\sum\left(\alpha - m_\alpha \right)^2}{\sum \left(\alpha\right)^2}\right)},
\end{align}
where the sums are taken over all pixels of the reconstruction. For the motivation behind this choice see Sec.~3 in Res2013. 

In Sec.~\ref{Ssec: mainres} we focus exclusively on the reconstruction of the two signals $s$ and $\alpha$. Then, in Sec.~\ref{Ssec: Compare}, we compare the results from multi-frequency \textsc{resolve} with standard imaging procedures. The reconstruction of the signal power spectra is discussed separately in Sec. \ref{Ssec: Pspec}. 

\subsection{Main test results}\label{Ssec: mainres}

We start by showing the reconstruction of the presented simulated observations using multi-frequency \textsc{resolve}. In Fig.~\ref{fig: recon}, a surface brightness and a spectral index signal are shown, together with the respective reconstructions obtained with \textsc{resolve} and absolute difference maps of both signals to their reconstructions. We show the spectral index maps in full, and over-layed with a mask that focuses on the part of the observed field that contains the brightest part of the surface brightness signal. Later, in Fig.~\ref{fig: comparison}, we show the reconstruction for different masks (along with a comparison to other methods, see Sec.~\ref{Ssec: Compare}). We choose the three different masks mainly qualitatively by visual comparison to show only the parts of the sky with a surface brightness signal above a certain threshold. The actual threshold values for the surface brightness were $2$ Jy/px, $4$ Jy/px, and $6$ Jy/px. In Fig.~\ref{fig: recon}, the second conservative mask is used. The error measures are $\delta_{I} = 0.13$ for the surface brightness, and $\delta_{\alpha} = 0.35$ for the spectral index.

It can be seen that \textsc{resolve} recovers very accurately the original surface brightness. For this particular choice of relative low noise, the algorithm succeeds in reconstructing even small scale features of the signal. The effects of the instrumental point spread function are successfully deconvolved. These findings are in perfect agreement with the results presented on single-frequency \textsc{resolve} in Res2013.

The general structure of the spectral index is well reconstructed, even for outer regions where the brightness signal is very weak. Overall, small scale features are much better recovered in the inner regions, where the main brightness sources are located, as can be seen by comparison between the full and the masked images in Fig.~\ref{fig: recon}. It is expected that the quality of the spectral index reconstruction depends on the strength of the observed surface brightness, as is illustrated by the poor performance of standard methods in recovering any structures outside the strong source regions, presented in the next section (see Fig.~\ref{fig: comparison}). In most real applications, the outer parts of the observed fields should therefore usually be not a focus of the investigation. We thus choose the relatively conservative mask in Fig.~\ref{fig: recon} to highlight this important part of the reconstruction. Nevertheless, it should be noted that -- at least for this low noise example with relatively high sensitivity due to the broad bandwidth -- \textsc{resolve} is able to reliably extrapolate its estimation also into the weaker regions around the main sources.

In addition, with \textsc{resolve}, an uncertainty of the spectral index reconstruction can be estimated. As for the single frequency reconstructions presented in Res2013, the second derivative of the posterior is used to approximate its covariance $D_\alpha$ (for details, see App.~\ref{App1} and Res2013). This way, the full estimate for the spectral index signal becomes

\begin{equation}
\alpha_x \approx (m_{\alpha})_x \pm \left(\sqrt{D_\alpha}\right)_{xx}. \label{fullestimate}
\end{equation}

For the given set of simulated data, roughly $60 \ \%$ of the original signal values within the unmasked regions in Fig.~\ref{fig: recon} lie within a $1-\sigma$ interval , but only roughly $20 \ \%$ lie within a $1-\sigma$ interval for the full spectral index image. Due to the non-linear nature of the inference problem (\ref{posmean2map}), it is expected that the uncertainty estimate is not exactly what would be expected from a pure Gaussian covariance (i.e. $68 \ \%$ in the $1-\sigma$ interval), the MAP estimate is not guaranteed to lie very close to the real posterior covariance (see Res2013 for this problem). It is no surprise that the estimate worsens for all the outer regions with only very weak brightness structures present, which explains the poor performance on the whole spectral index map. Furthermore, as discussed in Res2013, the calculation of an uncertainty estimate is computationally costly. Due to a lack of accessible computer power only for testing purposes, we stopped the calculations at some point and smoothed the outcome. This should not be a great problem, since the uncertainty is expected to be smooth and we are mainly interested in a proof of concept at this point. 

\subsection{Comparison to standard methods}\label{Ssec: Compare}

In this section, we compare the results of multi-frequency \textsc{resolve} with two standard methods: A straight forward power-law fit for (\ref{eq:basicspecdep}) using single-frequency CLEAN images, and a MF-MS-CLEAN reconstruction for the surface brightness and the spectral index. The reconstructions were performed on the same simulated observation as in Sec.~\ref{Ssec: mainres}. All results were obtained using the radio-astronomical software package CASA \citep{CASA}. The MF-MS-CLEAN reconstructions were obtained using 1500 iterations, a small gain factor of $0.1$, uniform weighting and ten different multi-scales ranging from a single pixel to moderately large structures. We further used two terms for the Taylor expansion in $\nu$ (see Sec.~\ref{SSec: MFSalpha}) and only half the available bandwidth, since a larger frequency range seems to lie outside the convergence radius of the Taylor-expansion (see Sec.~\ref{SSec: MFSalpha} used in the implementation of CASA.

In Fig.~\ref{fig: comparison}, a comparison is shown between \textsc{resolve}, a power-law fit, and MF-CLEAN results for the spectral index and differently strong masks. The full field is not shown, because neither the power-law fit nor the CLEAN reconstruction could recover any structures in the remaining regions. The error measures are listed in Tab.~\ref{tab: comparison}. It can be seen that for this example \textsc{resolve} overall outperforms the other methods. The advantage of \textsc{resolve} is more pronounced in the outer regions, where the surface brightness signal is weak, but \textsc{resolve} still gives the best result even for the most central parts of the reconstruction (see also the full reconstruction in Fig.~\ref{fig: recon}). 

Fig.~\ref{fig: brightcomparison} shows a comparison of multi-frequency surface brightness reconstructions with \textsc{resolve} and MS-MF-CLEAN for the simulated observation of Sec.~\ref{Ssec: mainres}. For this simulation, \textsc{resolve} is more successful in reconstructing the overall structure, and especially recovers more of the small scales. It should be noted that this particular CLEAN image was achieved using the full bandwidth and coverage (other than for the spectral index images, as stated at the beginning of the section).

\begin{table}
  \centering
  \footnotesize
  \tabcolsep=0.06cm
  \begin{tabular}{l|c|c|c}
  \hline
  \hline
  Algorithm \textbackslash Mask  & liberal mask & medium mask & conservative mask \\
  \hline
  \textsc{resolve} & \hspace{8px} 0.34 & \hspace{8px} 0.33 &\hspace{8px} 0.32 \\
  power-law fit & 132.79 &  \hspace{8px} 0.94 & \hspace{8px} 0.48\\
  MS-MF CLEAN & \hspace{8px} 3.07 & \hspace{8px} 1.89 & \hspace{8px} 1.84\\
  \hline
  \hline
  \end{tabular}
  \caption{$\mathcal{L}_2$ error measures for \textsc{resolve}, the power law fit, and MS-MF CLEAN for the simulation and the reconstruction observation of Sec.~\ref{Ssec: mainres} and the three different masks defined in Sec.~\ref{Ssec: mainres}.}
  \label{tab: comparison}
\end{table}

\subsection{Power spectrum reconstructions}\label{Ssec: Pspec}

The power spectrum of the spectral index is reconstructed together with the field itself. It represents the spatial correlation of the spectral index over the observed sky. As explained in Sec.~\ref{SSec: MFResolve}, we expect the typical spectral index structure of an extended (i.e. spatially correlated) radio source to be spatially correlated to itself. For all the details on power spectrum reconstructions we refer the reader to Res2013, since almost all derivations for standard single-frequency \textsc{resolve} power spectrum reconstructions are valid as well for the spectral index. 

In this section we only discuss the spectral index power spectrum, since the reconstructions for the brightness power spectra are identical to the ones already presented in Res2013. We simply note that, in principle, a multi-frequency reconstruction recovers structure on smaller scales, and thus, it is expected that multi-frequency \textsc{resolve} should be able to map spatial power spectra up to higher Fourier modes (i.e smaller correlation structures).

\begin{figure}[h!]
\centering
  \subfigure[Spectral power spectrum reconstruction]{
   \includegraphics[width=0.45\textwidth]{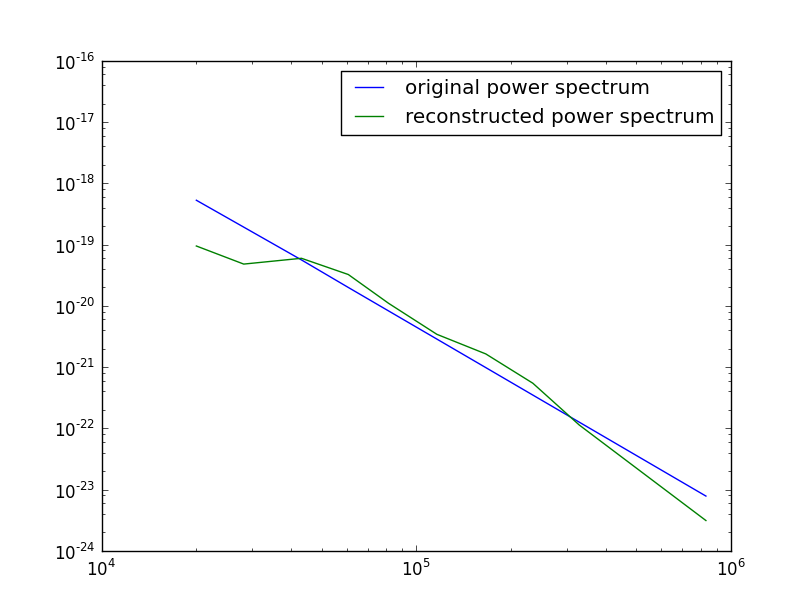}}
  \caption{Power spectrum reconstruction of the spectral index $\alpha$ for the reconstruction shown in Sec.~\ref{Ssec: mainres} using multi-frequency \textsc{resolve}.} 
\label{fig: pspec}
\end{figure}

A typical result from multi-frequency \textsc{resolve} is illustrated in Fig.~\ref{fig: pspec}. It shows the original spectral index power spectrum of the Gaussian signal field, used in the simulated observations of Sec.~\ref{Ssec: mainres}, and its reconstruction that belongs to the same iteration step as the presented signal reconstructions earlier. The power spectrum is reconstructed relatively well, no power is lost on high modes (i.e. small scales), which is just a consequence of the fact that the simulated observation was conducted with low noise.

As for the brightness reconstructions in Res2013, we emphasize that the power spectrum should not only be viewed as a by-product of the algorithm in order to accurately estimate the spectral index signal. For instance, some astronomical objects show very distinct spectral index structures and can even be classified after this criterion. A prominent example might be radio halos and relics of galaxy clusters, both of which typically show steep spectral indices that evolve spatially roughly like the source itself \citep{ClusterReview}. Measuring the spatial power spectra of these objects might lead to a more exact and quantitative classification scheme. Another application might lie in the investigation of different physical processes within a single source that lead to spectrally very different regions. Estimating the spectral correlation structure over such regions offers a new way of quantitative analysis of the interplay of these processes.   

\section{Conclusion} \label{Sec: Conclusion}

We presented a multi-frequency extension to the imaging algorithm \textsc{resolve} \citep{resolve}. The combined algorithm is optimal for multi-frequency imaging of extended radio sources. It simultaneously estimates the surface brightness at a reference frequency, and the spectral index of the source. Within the assumption of a spectral index model, no further expansions or parameter-dependent modeling is used in the reconstruction. Multi-frequency \textsc{resolve} is thus capable of exploiting the full bandwidth of a modern radio observation for maximum sensitivity and resolution, only limited by higher order spectral effects like spectral curvature. 

Multi-frequency \textsc{resolve} has been tested successfully using simulated observations of the VLA in its A-configuration. For the presented tests, the algorithm can outperform standard imaging methods in both surface brightness and spectral index estimations.

The algorithm uses a Gaussian prior and an effective log-normal model for the spectral index. This approach is not necessarily restricted to a combination with single-frequency \textsc{resolve}, and can in principle be combined with any other imaging or deconvolution method for the surface brightness reconstruction.

For the sake of feasibility, many details of radio interferometric observations have been left out of the analysis. The response might realistically contain a number of additional effects, for instance an instrumental primary beam or wide-field and direction dependent effects. We refer the reader to \citet{resolve}, where many of these problems already have been discussed in the outlook.

In its current form, the presented algorithm has relatively high computational costs and numerical demands \citep[for an analysis of algorithmic efficiency, see][]{resolve}. In general, this is true for most Bayesian statistical inference algorithms. This could pose an obstacle for day-to-day applicability, since especially modern broad band data sets tend to be very large, and therefore are already a numerically challenge on their own. Since this study was centered on fundamental algorithmic development, numerical efficiency was not a focus. Future work might be needed to obtain a more efficient implementation of the algorithm.

For the future, it seems to be desirable to refrain completely from an explicit spectral model, and try to infer a full, non-parametric, three dimensional spectral intensity $I(l,m,\nu)$. Possibly, such a development could benefit greatly from reconstructing a full cross-correlation structure between the sky and spectral space. We leave this more complete approach for a future publication.  
  
\bibliographystyle{aa}
\bibliography{references.bib}

\appendix

\section{Details of the algorithm} \label{App1}

In this appendix, we repeat some of the basic derivations from Res2013, and derive the details of how reconstructing the spectral index using \textsc{resolve} differs from the standard procedure presented there. In principle, many of the original derivations are still valid for multi-frequency \textsc{resolve}, estimating $s$ and $\alpha$. The derivations for power spectrum reconstructions and uncertainty calculations stay especially close to the equations already presented in Res2013, and we do not repeat them explicitly here. 

\subsection{Reconstruction of the sky brightness signal field $s$}\label{App1:1}

The estimation of the sky brightness at a reference frequency $I_{0}(l,m,\nu_0) = \rho_0 \mathrm{e}^{s(l,m)}$ can mostly be conducted with the standard single-frequency \textsc{resolve}. 

The only difference is the more complex response operator $R_{(\mathrm{s})} = W(k,\nu) \mathcal{F} \left(\mathrm{e}^{-\ln{\left(\nu/\nu_0\right)} \alpha} \ \circ \right)$. Dropping the explicit signal index for a moment and writing the specific measurement model (\ref{mfresolve}) for the signal $s$ with explicit operator and field indices

\begin{align}
d_{k\nu} = R_{k\nu x} \ (\rho_0 \ \mathrm{e}^{s})_x + n_{k\nu}, \label{sdmodel}
\end{align}
reveals that the operator $R_{k\nu x}$ actually spans the single-frequency Fourier transform over all observed frequencies into a large, many-frequency data space\footnote{As a reminder: In (\ref{sdmodel}) we use an implicit convention to sum or taking the integral over repeated discrete or continuous indices, see Res2013 for details.}. Conversely, the adjoint response $R^\dagger$ now includes a sum over all frequencies to collapse everything back into the two dimensional sky at reference frequency. Since both operations are used in \textsc{resolve}, it is by this procedure that the multi-frequency \textsc{resolve} effectively uses $uv$-information at all frequencies to constrain the estimation of $s$. Of course, the quality of this reconstruction depends on the accuracy of the current estimation for $\alpha$, used in the reconstruction step for $s$ to define $R_{(\mathrm{s})}$.

\subsection{Reconstruction of the spectral index $\alpha$}

The log-normal model for the spectral index, $\mathrm{e}^{-\ln{\left(\nu/\nu_0\right)} \alpha}$, only differs in the term $b=-\ln{\left(\nu/\nu_0\right)}$ from the model used for the sky brightness signal. This slightly complicates the calculations for the signal estimation with MAP and the power spectrum reconstruction and eventual uncertainty calculation. For both, derivatives of the posterior (\ref{a_posterior}) with respect to $\alpha$ are needed (see Res2013), and this will add extra $b$-terms into the equations. The results of this are summarized in Sec.~\ref{Ssec: Fullcomb}. 

As for the sky brightness signal $s$ (see App.~\ref{App1:1}), the response operator for the $\alpha$ reconstructions $R_{(\mathrm{\alpha)}} =  W(k,\nu) \mathcal{F} \left(\mathrm{e}^{s(l,m)} \ \circ \right)$ is more complex. Effectively, the $b$-term in the basic log-normal model  $\mathrm{e}^{-\ln{\left(\nu/\nu_0\right)} \alpha}$ would prevent us from just defining our signal space to be the two-dimensional sky since $R_{(\mathrm{\alpha)}}$ acts on $\mathrm{e}^{-\ln{\left(\nu/\nu_0\right)} \alpha}$. In the actual implementation of the multi-frequency \textsc{resolve} algorithm, we circumvent this problem by assuming that $R_{(\mathrm{\alpha)}}$ acts on $\alpha$ and all other terms, including the exponential operation of the log-normal model, are part of the operator\footnote{This actually renders $R_{(\mathrm{\alpha)}}$ a non-linear operator.}. In this way, $R_{(\mathrm{\alpha)}}$ and $R_{(\mathrm{\alpha)}}^\dagger$ can be understood to act in the same way as their respective counterparts for $s$ do (i.e. they also span up and collapse into the full frequency data space).

\subsection{Combined algorithm}\label{Ssec: Fullcomb}

Using our findings in Res2013 and in the previous subsections, multi-frequency \textsc{resolve} comes down to solving iteratively two only slightly different, subsequent sets of equations:

\begin{align}
&\mathrm{\underline{Estimation \ of \ s}} \notag\\
&S^{-1}_{p} \ m_{\mathrm{s}} + \mathrm{e}^{m_{\mathrm{s}}} \cdot M_{\mathrm{s}} \mathrm{e}^{m_{\mathrm{s}}} - j_{\mathrm{s}} \cdot \mathrm{e}^{m_{\mathrm{s}}} = 0 \label{eq:one}\\ 
\notag\\
&\left(D_{\mathrm{s}}\right)_{xy} = S^{-1}_{p \ xy} +  \mathrm{e}^{(m_{\mathrm{s}})_x} \ (M_{\mathrm{s}})_{xy} \ \mathrm{e}^{(m_{\mathrm{s}})_y} \notag\\
& \ \ \ \ \ \ \ \ \ + \mathrm{e}^{(m_{\mathrm{s}})_y} \int dz \ M_{\mathrm{s}}(x,z) \ \mathrm{e}^{(m_{\mathrm{s}})_z} \notag\\
& \ \ \ \ \ \ \ \ \ - (j_{\mathrm{s}})_x \cdot \mathrm{e}^{(m_{\mathrm{s}})_x} \ \delta_{xy} \label{eq:two}\\
\notag\\
&p_i^s = \frac{\left(q_i^s + \frac{1}{2} \mathrm{tr} \left((m_{\mathrm{s}}{m_{\mathrm{s}}}^{\dagger} + D_{\mathrm{s}})S^{(i)}\right)\right)}{\left(\alpha_i^{(\mathrm{s-pr})} - 1 + \frac{\varrho_i}{2} + (Tp)_i\right)} \label{eq:three} \\
\notag\\
\notag\\
&\mathrm{\underline{Estimation \ of \ \alpha}} \notag\\
&A^{-1}_{p} m_{\mathrm{\alpha}} + b \ \mathrm{e}^{b m_{\mathrm{\alpha}}} \cdot M_{\mathrm{\alpha}} \mathrm{e}^{b m_{\mathrm{\alpha}}} - j_{\mathrm{\alpha}} \cdot b \ \mathrm{e}^{b m_{\mathrm{\alpha}}} = 0 \label{eq:estimate_a one}\\ 
\notag\\
&\left(D_{\mathrm{\alpha}}\right)_{xy} = A^{-1}_{p \ xy} +  b^2 \ \mathrm{e}^{b (m_{\mathrm{\alpha}})_x} \ (M_{\alpha})_{xy} \ \mathrm{e}^{b (m_{\mathrm{\alpha}})_y} \notag\\
& \ \ \ \ \ \ \ \ \ + b^2 \ \mathrm{e}^{b (m_{\mathrm{\alpha}})_x} \int dz \ M_{\mathrm{\alpha}}(x,z) \ \mathrm{e}^{b (m_{\mathrm{\alpha}})_z} \notag\\
& \ \ \ \ \ \ \ \ \ - (j_{\mathrm{\alpha}})_x \cdot \mathrm{e}^{(m_{\mathrm{\alpha}})_x} \ \delta_{xy} \label{eq:estimate_a two}\\
\notag\\
&p_i^\alpha = \frac{\left(q_i^\alpha + \frac{1}{2} \mathrm{tr} \left((m_{\mathrm{\alpha}}{m_{\mathrm{\alpha}}}^{\dagger} + D_{\mathrm{\alpha}})A^{(i)}\right)\right)}{\left(\alpha_i^{(\mathrm{\alpha-pr})} - 1 + \frac{\varrho_i}{2} + (Tp)_i\right)} \label{eq:estimate_a three}
\end{align}
A rigorous derivation for all equations can be found in Res2013. The two sets of equations only differ in form by the $b=-\ln{\left(\nu/\nu_0\right)}$ - terms that show up in the spectral index reconstruction because of the derivatives used in order to calculate the MAP estimate. The quantities $j_{\mathrm{s}}$, $M_{\mathrm{s}}$  and $j_{\mathrm{\alpha}}$, $M_{\mathrm{\alpha}}$ are defined as

\begin{align}
j_{\mathrm{s}} &= R_{\mathrm{s}}^{\dagger}N^{-1}d, \\
M_{\mathrm{s}} &= R_{\mathrm{s}}^{\dagger}N^{-1} R_{\mathrm{s}}, \\
j_{\mathrm{\alpha}} &= R_{\mathrm{\alpha}}^{\dagger}N^{-1}d, \\
M_{\mathrm{\alpha}} &= R_{\mathrm{\alpha}}^{\dagger}N^{-1} R_{\mathrm{\alpha}}.
\end{align}
$S^{(i)}$, or $A^{(i)}$ are projection operators onto a band of Fourier modes denoted by the index $i$, while $p_i$ are parameters to model the unknown power spectrum into a number of such bands $S = \sum_{i} p_i S^{(i)}$, or  $A = \sum_{i} p_i A^{(i)}$ (see Res2013 for details). The quantities $q$, $\alpha^{(\mathrm{pr})}$ and $\varrho$ are parameters of a power spectrum prior for the signal or the spectral index, and $T$ is an operator, which enforces a smooth solution of the power spectrum $p_i$.     

Eqs. (\ref{eq:one}) and (\ref{eq:estimate_a one}) are the fix point equations that need to be solved numerically to find a MAP signal estimate $m_{\mathrm{s}}$ or $m_{\mathrm{\alpha}}$ for the current iteration. The second equations (\ref{eq:two}) and (\ref{eq:estimate_a two}) result from calculating the second derivative of the respective posteriors for the signal estimates $m_{\mathrm{s}}$ or $m_{\mathrm{\alpha}}$, their inverses serve as an approximation to the signal uncertainty $D_{\mathrm{s}} = \langle (s-m_s) (s-m_s)^\dagger \rangle$ or $D_{\mathrm{\alpha}} = \langle (\alpha-m_\alpha) (\alpha-m_\alpha)^\dagger \rangle$ at each iteration step. The last equations (\ref{eq:three}) and (\ref{eq:estimate_a three}) represent an estimate for the signal power spectra (and therefore their autocorrelation functions), using the signal uncertainties $D_{\mathrm{s}}$ or $D_{\mathrm{\alpha}}$ to correct for missing signal power in the current estimates $m_{\mathrm{s}}$ or $m_{\mathrm{\alpha}}$ . The iteration is stopped after a suitable convergence criterion is met (see App.~2 in Res2013).

\begin{figure*}[p]
   \centering
	\subfigure[Surface brightness signal $\mathrm{e}^s$.]{
		\includegraphics[width=0.30\textwidth]{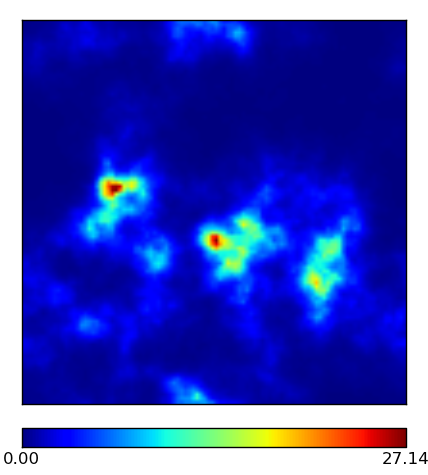}}
	\subfigure[Masked spectral index signal $\alpha$]{
		\includegraphics[width=0.30\textwidth]{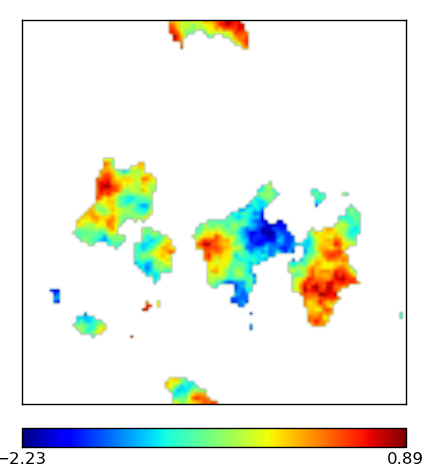}}
	\subfigure[Spectral index signal $\alpha$]{
		\includegraphics[width=0.30\textwidth]{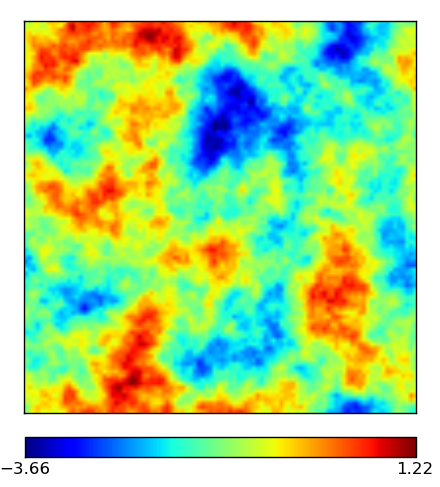}}\\
        \subfigure[\textsc{resolve} surface brightness reconstruction $\mathrm{e}^{m_{s}}$.]{
		\includegraphics[width=0.30\textwidth]{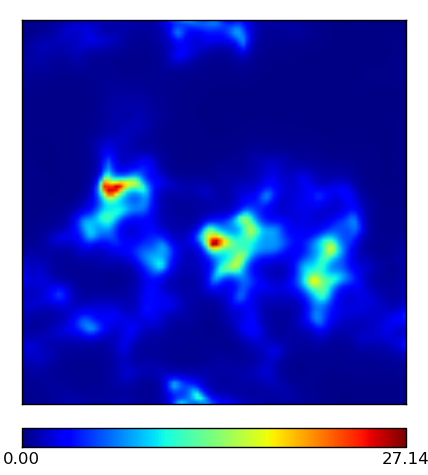}}
	\subfigure[Masked \textsc{resolve} spectral index reconstruction $m_\alpha$.]{
		\includegraphics[width=0.30\textwidth]{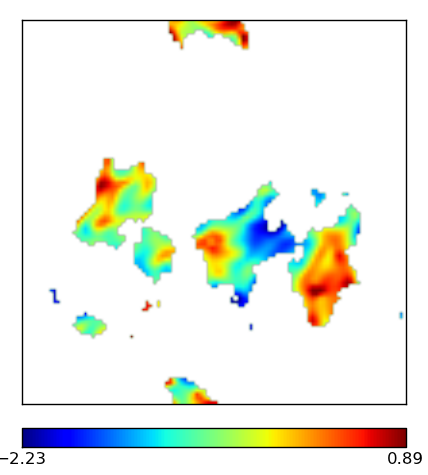}}
	\subfigure[\textsc{resolve} spectral index reconstruction $m_\alpha$.]{
		\includegraphics[width=0.30\textwidth]{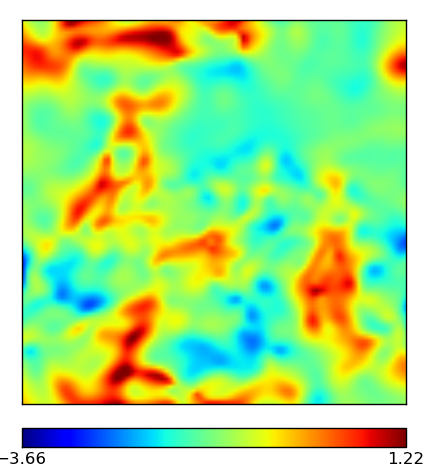}}\\ 
	\subfigure[Absolute error $\left|\mathrm{e}^s-\mathrm{e}^{m_{s}}\right|$.]{
		\includegraphics[width=0.30\textwidth]{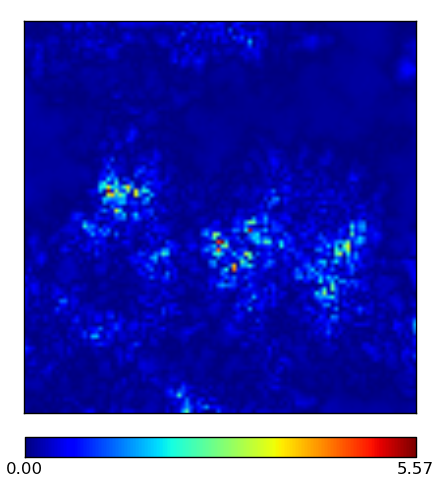}}
	\subfigure[Masked absolute error $\left|\alpha-m_{\alpha}\right|$.]{
		\includegraphics[width=0.30\textwidth]{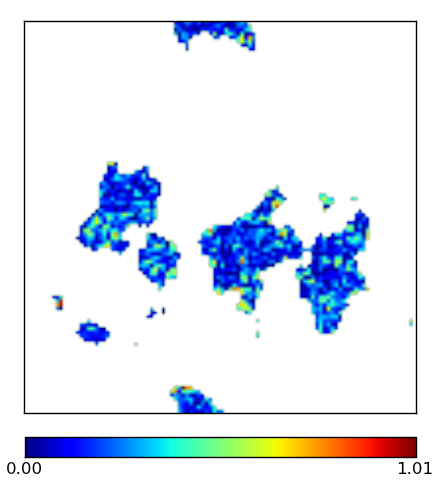}}
	\subfigure[Absolute error $\left|\alpha-m_{\alpha}\right|$.]{
		\includegraphics[width=0.30\textwidth]{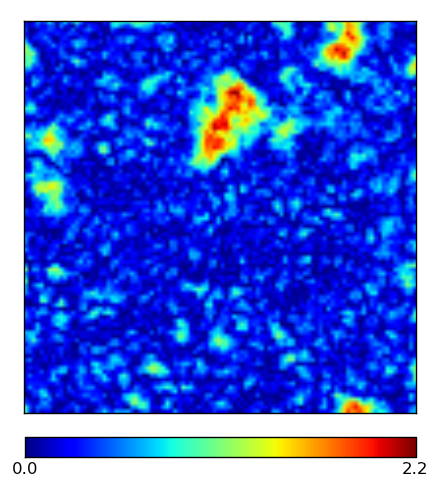}}\\
	\caption{Multi-frequency reconstruction of the two signal fields $\mathrm{e}^s$ and $\alpha$, observed with a sparse $uv$-coverage from a VLA-A-configuration (see Fig.~\ref{realuvcov}). The images are $100^2$ pixels large, the pixel size corresponds to roughly $0.1$ arcsec. The brightness units are in Jy/px. The ridge-like structures  in the difference maps simply stem from taking the absolute value and mark zero-crossings between positive and negative errors.} 
\label{fig: recon}  
\end{figure*}

\begin{figure*}[p]
   \centering
        \subfigure[Signal]{
		\includegraphics[width=0.23\textwidth]{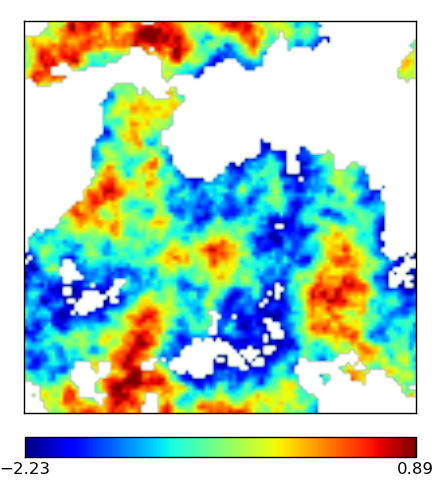}}
	\subfigure[\textsc{resolve}]{
		\includegraphics[width=0.23\textwidth]{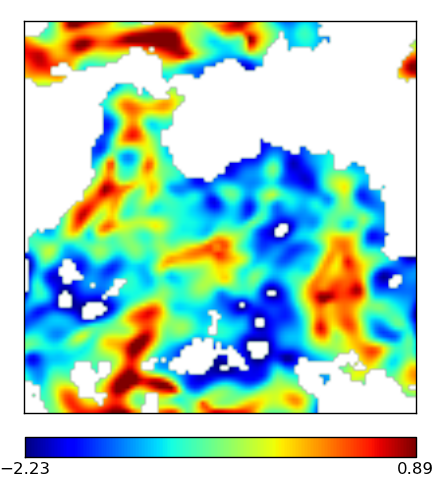}}
	\subfigure[Power-law fit]{
		\includegraphics[width=0.23\textwidth]{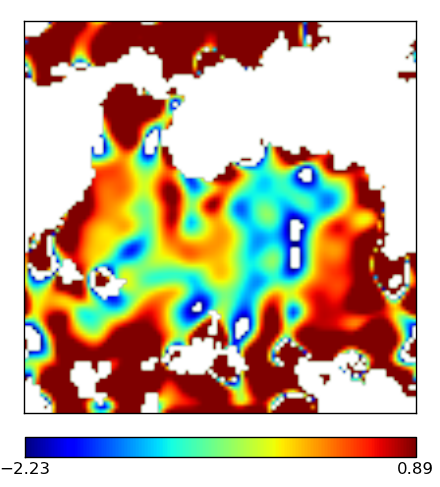}}
	\subfigure[MS-MF-CLEAN]{
		\includegraphics[width=0.23\textwidth]{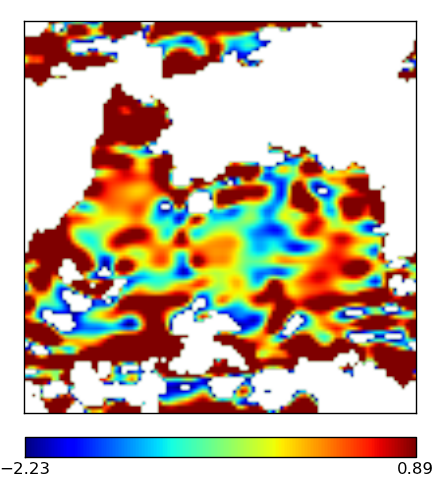}}\\
	\subfigure[Signal]{
		\includegraphics[width=0.23\textwidth]{aweakmask_crop.png}}
	\subfigure[\textsc{resolve}]{
		\includegraphics[width=0.23\textwidth]{maweakmask_crop.png}}
	\subfigure[Power-law fit]{
		\includegraphics[width=0.23\textwidth]{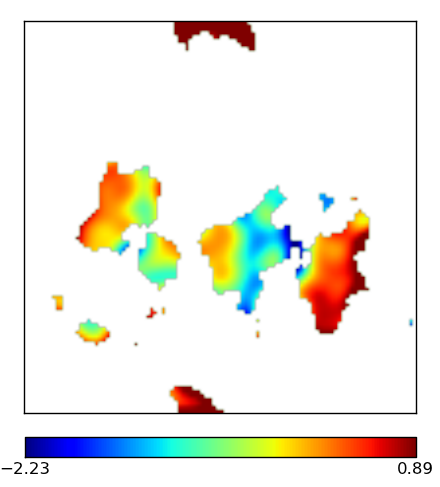}}
	\subfigure[MS-MF-CLEAN]{
		\includegraphics[width=0.23\textwidth]{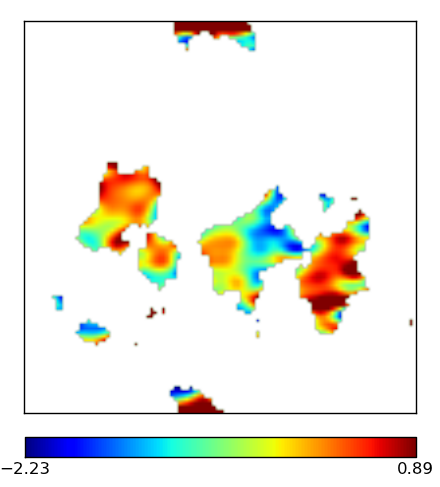}}\\
	\subfigure[Signal]{
		\includegraphics[width=0.23\textwidth]{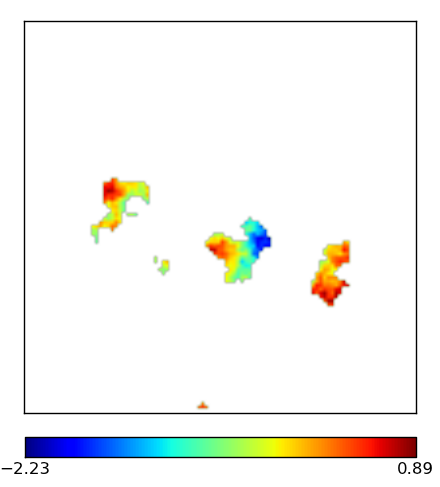}}
	\subfigure[\textsc{resolve}]{
		\includegraphics[width=0.23\textwidth]{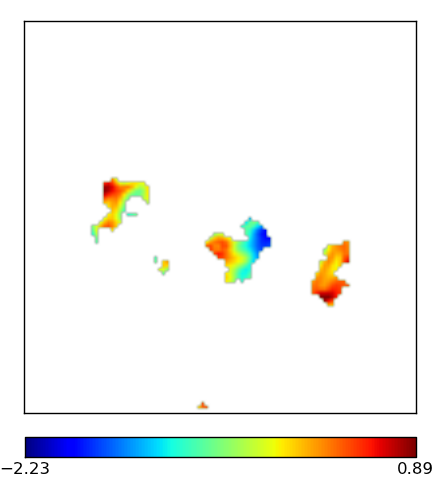}}
	\subfigure[Power-law fit]{
		\includegraphics[width=0.23\textwidth]{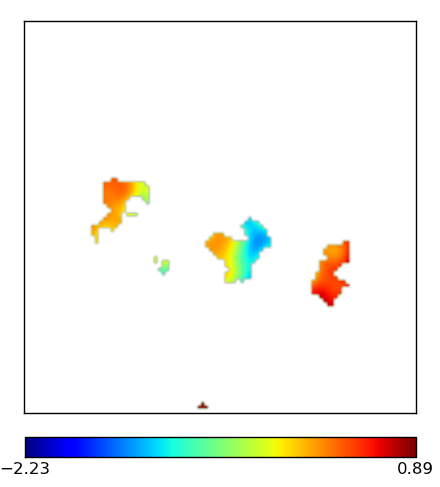}}
        \subfigure[MS-MF-CLEAN]{
		\includegraphics[width=0.23\textwidth]{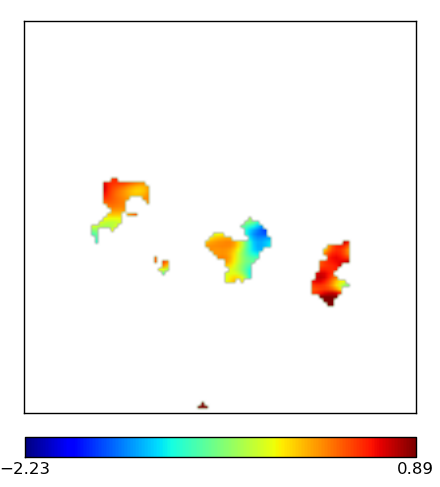}}\\
	\caption{Comparison of different methods for spectral index reconstruction with differently strong masks. The images are $100^2$ pixels large, the pixel size corresponds to roughly $0.1$ arcsec. \textit{First column}: Signal. \textit{Second column}: \textsc{resolve} reconstruction. \textit{Third column} Power-law fit. \textit{Fourth column}: MS-MF-CLEAN.}
\label{fig: comparison}
\end{figure*}

\begin{figure*}[p]
   \centering
	\subfigure[\textsc{resolve} brightness reconstruction $\mathrm{e}^{m_s}$]{
		\includegraphics[width=0.44\textwidth]{m_crop.png}}
	\subfigure[MS-MF CLEAN brightness reconstruction]{
		\includegraphics[width=0.44\textwidth]{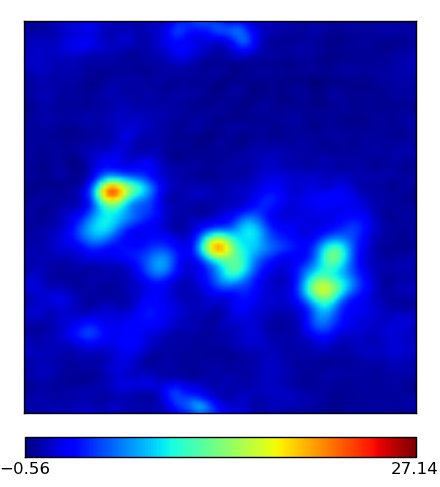}} \\
        \subfigure[Absolute error $\left|\mathrm{e}^s-\mathrm{e}^{m_{s}}\right|$]{
		\includegraphics[width=0.44\textwidth]{m_absdiff_crop.png}}
	\subfigure[Absolute error $\left|\mathrm{e}^s-m_{\mathrm{clean}}\right|$]{
		\includegraphics[width=0.44\textwidth]{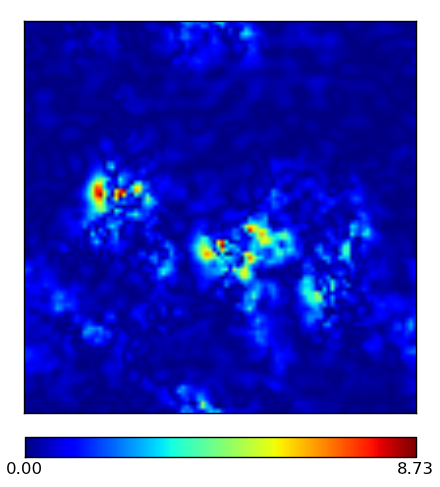}} \\
	\caption{Comparison of multi-frequency \textsc{resolve} and MS-MF-CLEAN surface brightness reconstructions. The images are $100^2$ pixels large, the pixel size corresponds to roughly $0.2$ arcsec. The brightness units are in Jy/px. The ridge-like structures in the difference maps simply stem from taking the absolute value and mark zero-crossings between positive and negative errors. \textit{First row left}: \textsc{resolve} reconstruction . \textit{First row right}: MS-MF CLEAN reconstruction. \textit{Second row left} Absolute per-pixel difference between the signal and the \textsc{resolve} reconstruction \textit{Second row right}: Absolute per-pixel difference between the signal and the MS-MF CLEAN reconstruction.}
  \label{fig: brightcomparison}
\end{figure*}

\end{document}